\documentclass[12pt]{article}
\pdfoutput=1
\usepackage{appendix}
\usepackage{anysize}
\usepackage{graphics}
\usepackage{color}
\usepackage{amssymb}
\usepackage{graphicx}
\usepackage{epstopdf}
\usepackage[hang,small,bf]{caption}
\usepackage{hyperref}
\usepackage{amsmath}
\usepackage{latexsym}
\usepackage{setspace}
\usepackage{sectsty}
\usepackage{breqn}
 
\usepackage{graphicx}
\usepackage{dcolumn}
\usepackage{bm}
\usepackage{epstopdf}
\usepackage{color}

\newcommand{\be}{\begin{equation}}
\newcommand{\ee}{\end{equation}}
\newcommand{\bea}{\begin{eqnarray}}
\newcommand{\eea}{\end{eqnarray}}

\def\4vol{{\int d^4x \sqrt{-g}}}

\def\beq{\begin{equation}}
\def\eeq{\end{equation}}
\def\bea{\begin{eqnarray}}
\def\eea{\end{eqnarray}}
\def\bitem{\begin{itemize}}
\def\eitem{\end{itemize}}

\newcommand{\nc}{\newcommand}
\nc{\nt}{\tilde{N}}
\nc{\ra}{\rightarrow}
\nc{\lsim}{\begin{array}{c}\,\sim\vspace{-21pt}\\< \end{array}}
\nc{\gsim}{\begin{array}{c}\sim\vspace{-21pt}\\> \end{array}}
\nc{\tnt}{\tilde{N}}
\nc{\tst}{\tilde{t}}

\nc{\LL}{L}
\nc{\vv}{\tilde{v}}
\title{
\begin{flushright}
\normalsize{
ANL-HEP-PR 88\\ EFI 11-35\\ FERMILAB-PUB-11-659-T
}
\end{flushright}
\vspace*{5mm} \Large\textbf{A 125 GeV SM-like Higgs in the MSSM and the $\gamma \gamma$ rate}
\vspace*{1.0cm}
\author{\textbf{Marcela Carena$^{a,b}$, Stefania Gori$^{a,c}$ } \\
\textbf{ Nausheen R.~Shah$^b$, Carlos E.~M.~Wagner$^{a,c,d}$ } \\
~\\
\normalsize\emph{$^a$Enrico Fermi Institute, University of Chicago, Chicago, IL 60637} \\
\normalsize\emph{$^b$Fermi National Accelerator Laboratory, P.~O.~Box 500, Batavia, IL 60510, USA}\\
\normalsize\emph{$^c$HEP Division, Argonne National Laboratory, 9700 Cass Ave., Argonne, IL 60439}\\
\normalsize\emph{$^c$Kavli Institute for Cosmological Physics, University of Chicago, Chicago, IL 60637}
}
}
\begin{document}

%\date{\today}
\setcounter{page}{0}
\maketitle
%\begin{center}
\begin{abstract}
We consider the possibility of a Standard Model~(SM)-like Higgs in the context of the Minimal Supersymmetric Standard Model~(MSSM), with a mass of about 125~GeV and with a production times decay rate into two photons which is similar or somewhat larger than the SM one. The relatively large value of the SM-like Higgs mass demands stops in the several hundred GeV mass range with somewhat large mixing, or a large hierarchy between the two stop masses in the case that one of the two stops is light. We find that, in general, if the heaviest stop mass is smaller than a few TeV, the rate of gluon fusion production of Higgs bosons decaying into two photons tends to be somewhat suppressed with respect to the SM one in this region of parameters. However, we show that an enhancement of the photon decay rate may be obtained for light third generation sleptons with large mixing, which can be naturally obtained for large values of $\tan\beta$ and sizable values of the Higgsino mass parameter.
\end{abstract}
\thispagestyle{empty}
\newpage
\setcounter{page}{1}

%%%%%%%%%%%%%%%%%%%%%%%%%%%%%%%%
\section{Introduction}
\label{sec:Intro}
%%%%%%%%%%%%%%%%%%%%%%%%%%%%%%%%
The minimal supersymmetric extension of the Standard Model~(MSSM) provides a well motivated framework that is currently being tested at high energy colliders. Supersymmetry breaking introduces tens of free parameters which can only be fixed by comparison with experimental data~\cite{reviews}. One of the most solid predictions of the model is the presence of a relatively light Standard Model~(SM)-like Higgs boson~\cite{Higgs:1964pi},\cite{Higgs:1966ev} with a mass of the order of the weak scale. The precise value of this Higgs mass is strongly dependent on loop corrections which depend quartically on the top quark mass and logarithmically on the scale of the stop masses. For both the stop masses at the TeV scale, there is a maximal value for the SM-like Higgs mass, which has been computed at the one and two-loop level by different methods, and is about 130~GeV~\cite{Okada:1990vk}--\cite{Degrassi:2002fi}.

%\cite{Sung:2011bd},\cite{ATLAS,CMS}
For these reasons, the observation of a Higgs boson with SM-like properties and a mass above 130~GeV would put a very strong constraint on the realization of the MSSM. However, the Tevatron and LHC experiments have currently ruled out the presence of a SM-like Higgs boson above $\sim 130$~GeV at the 95\% confidence level~\cite{Aaltonen:2011gs}--\cite{CMS}, essentially ruling out most of the region of parameters which would be inconsistent with this model for supersymmetric particle masses in the TeV scale. In addition, the ATLAS and CMS experiments have reported signatures consistent with the presence of a Higgs particle with a mass of about 125~GeV and with a central value for the associated  photon rate that is similar or somewhat above the SM one~\cite{ATLAS,CMS}. The observed production rates in the $WW$ and $ZZ$ channels are consistent with those expected for a SM Higgs in that mass range.

Although it is premature to interpret the signatures observed at the LHC as evidence of Higgs production, we shall entertain the possibility that indeed they are associated with the presence of a Higgs with a mass of about 125~GeV, a photon production rate greater than or similar to the SM one for the same Higgs mass and at the same time a SM like production rate in the other channels~($WW$ and $ZZ$). In section 2 we shall discuss the Higgs mass predictions and the constraints that can be derived on the stop spectrum from the Higgs mass determination. In section 3 we shall discuss the Higgs production and decay rates. We shall analyze the rate of photon and weak boson production and their dependence on the values of the CP-odd Higgs mass. We further show that the slepton spectrum may have an important impact on the photon decay rate. In particular in the presence of large mixing in the third generation slepton sector, photon decay branching ratios larger than in the SM may be achieved for large values of the CP-odd Higgs mass. We shall reserve section 4 for our conclusions.

%%%%%%%%%%%%%%%%%%%%%%%%%%%%%%%%
\section{Higgs Mass Predictions}
\label{sec:HMass}
%%%%%%%%%%%%%%%%%%%%%%%%%%%%%%%%

As we have mentioned in the introduction, for almost degenerate stop masses, the Higgs mass depends logarithmically on the averaged stop mass scale, $M_{\rm SUSY}$. There is also a quadratic and quartic one-loop dependence on the stop mixing parameter, $\tilde{A}_t$. Specifically, for moderate or large values of $\tan\beta$, which is the ratio of the Higgs vacuum expectation values, and large values of the non-standard Higgs masses, characterized by the CP-odd Higgs mass, $m_A$, one gets~\cite{mhiggsRG1}
\begin{eqnarray}
m_h^2& \simeq & M_Z^2\cos^2 2\beta+\frac{3}{4\pi^2}\frac{m_t^4}{v^2}\left[ \frac{1}{2}\tilde{X}_t + t
+\frac{1}{16\pi^2}\left(\frac{3}{2}\frac{m_t^2}{v^2}-32\pi\alpha_3
\right)\left(\tilde{X}_t t+t^2\right) \right]\,, \label{mhsm}
\end{eqnarray}
where
\be
\label{escala}
t=\log\frac{ M_{\rm SUSY}^2}{m_t^2}\;.
\ee
The parameter $\tilde{X_t}$ is given by
\begin{eqnarray}
\label{stopmix}
\tilde{X}_{t} & =& \frac{2 \tilde{A}_t^2}{M_{\rm SUSY}^2}
                  \left(1 - \frac{\tilde{A}_t^2}{12 M_{\rm SUSY}^2} \right)\;,
\nonumber \\
\tilde{A}_t & = & A_t-\mu\cot\beta\;,
\end{eqnarray}
where $A_t$ is the trilinear Higgs-stop coupling and $\mu$ is the Higgsino mass parameter.

The above expression is only valid for relatively small values of the splitting of the stop masses. For larger splittings between the two stop soft masses, similar expressions may be found, for instance, in Refs.~\cite{mhiggsRG1}--\cite{Degrassi:2002fi}. Eq.~(\ref{mhsm}) has a maximum at large values of $\tan\beta$ and $A_t \simeq 2.4 M_{\rm SUSY}$ in the $\bar{DR}$ scheme, and as claimed in the introduction, gives $m_h\sim $ 130 GeV for a top quark mass of about 173~GeV and $M_{\rm SUSY}$ of the order of 1~TeV. The Higgs mass expression in Eq.~(\ref{mhsm}) is modified by thresholds effects on the top-quark Yukawa coupling, which depend on the product of the gluino mass and $A_t$, and which induce a small asymmetry in the Higgs mass expression with respect to the sign of $A_t$, leading to slightly larger values for positive $A_t M_3$ ~\cite{Carena:2000dp}.

There are additional contributions to Eq.~(\ref{mhsm}) that come from the sbottom and slepton sectors which can be important at large values of $\tan\beta$. The sbottom corrections are always negative and are given by
\begin{equation}
\Delta m_h^2 \simeq - \frac{h_b^4 v^2}{16 \pi^2 } \frac{\mu^4}{M_{\rm SUSY}^4 }\left( 1 + \frac{t}{16\pi^2}(9 h_b^2 - 5 \frac{m_t^2}{v^2} - 64 \pi  \alpha_3 )\right)\;,
\end{equation}
where the bottom Yukawa coupling $h_b$ is given by
\begin{equation}
h_b \simeq \frac{m_b }{v\cos\beta(1 + \tan\beta \Delta h_b)}\;,
\end{equation}
and $\Delta h_b$ is a one-loop correction whose dominant contribution depends on the sign of $\mu M_3$~\cite{deltamb,deltamb1,deltamb2}. Positive values of $\mu M_3$ tend to reduce the Yukawa coupling which therefore reduces the negative sbottom effect on the Higgs mass, while negative values of $\mu M_3$ enhance the Yukawa coupling and may diminish the Higgs mass for large values of $\tan\beta$.

Similarly, the corrections from the slepton sector are,
\begin{equation}\label{eq:mhstau}
\Delta m_h^2 \simeq - \frac{h_\tau^4 v^2}{48 \pi^2 } \frac{\mu^4}{M_{\tilde{\tau}}^4 }\,,
\end{equation}
where $M_{\tilde{\tau}}$ has been identified with the characteristic stau spectrum scale and we have ignored the logarithmic loop corrections. The $\tau$ Yukawa coupling, $h_\tau$, is given by
\begin{equation}
h_\tau \simeq \frac{m_\tau}{v\cos\beta(1 + \tan\beta \Delta h_\tau)}\;,
\end{equation}
and $\Delta h_\tau$ is a loop correction factor that depends on the sign of $\mu M_2$~\cite{deltamb2}\footnote{Positive values of $\mu M_2$ are preferred in order to reconcile the theoretical prediction for the muon anomalous magnetic moment with its experimental value\cite{reviews}.}.

From Eq.~(\ref{mhsm}) it follows that Higgs masses of about 125 GeV may only be obtained for values of the stop masses of the order of several hundred GeV and sizable values of $\tilde{A}_t > M_{\rm SUSY}$. The scale $M_{\rm SUSY}$ and/or the mixing parameter $\tilde{A}_t$ should take larger values if there is a significant negative sbottom or stau induced effect on the Higgs mass, which is possible for very large values of $\tan\beta$. We have used the program FeynHiggs~\cite{FH,Heinemeyer:1998yj,Heinemeyer:1998np,Frank:2006yh}~\footnote{Note that in FeynHiggs the $\Delta h_\tau$ corrections are not implemented. However, since these corrections can always be compensated for by a small modification of the values of $\mu$ and $A_\tau$, we do not expect that the introduction of these loop corrections will modify our results in the parameter space of interest.} for the computation of the Higgs cross sections~\cite{Spira:1995rr,Maltoni,Grazzini} and properties. We always compare our results with CPsuperH~\cite{CPsuperH}, which gives good agreement with the results of FeynHiggs, apart from the large $\tan\beta$ region, where stau effects, which are not included in CPsuperH, become significant. In order to determine the region of stop masses consistent with the Higgs signature~\cite{Carena:1998wq}, we have considered an uncertainty in the calculation of the Higgs masses of about 2~GeV, and hence, we conclude that the entire range of calculated Higgs masses between 123~GeV and 127~GeV may be consistent with the observed Higgs signatures.

\begin{figure}
\begin{center}
\begin{tabular}{cc}
(a)&(b)\\
&\\
\includegraphics[width=0.5\textwidth]{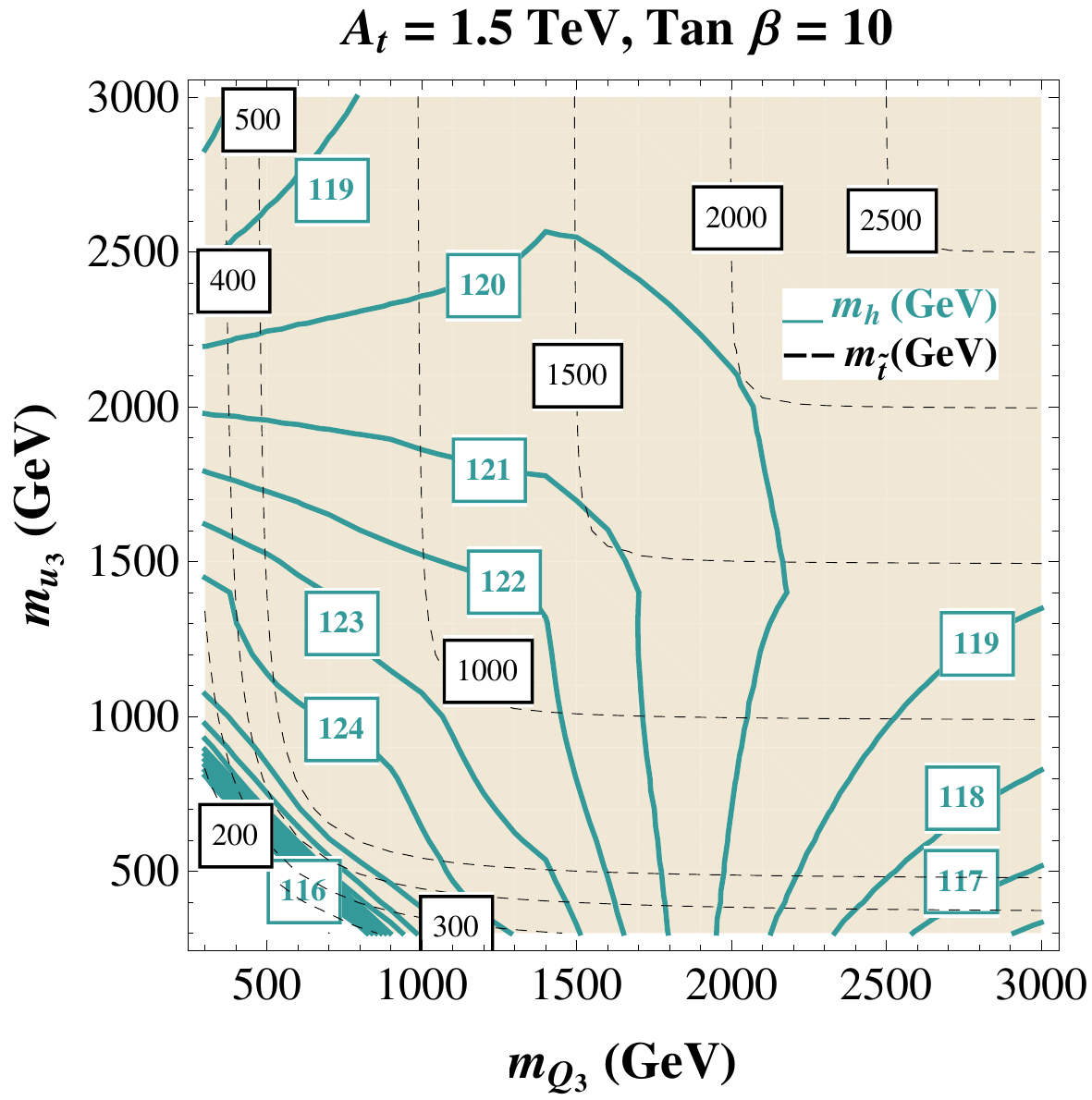} &
\includegraphics[width=0.5\textwidth]{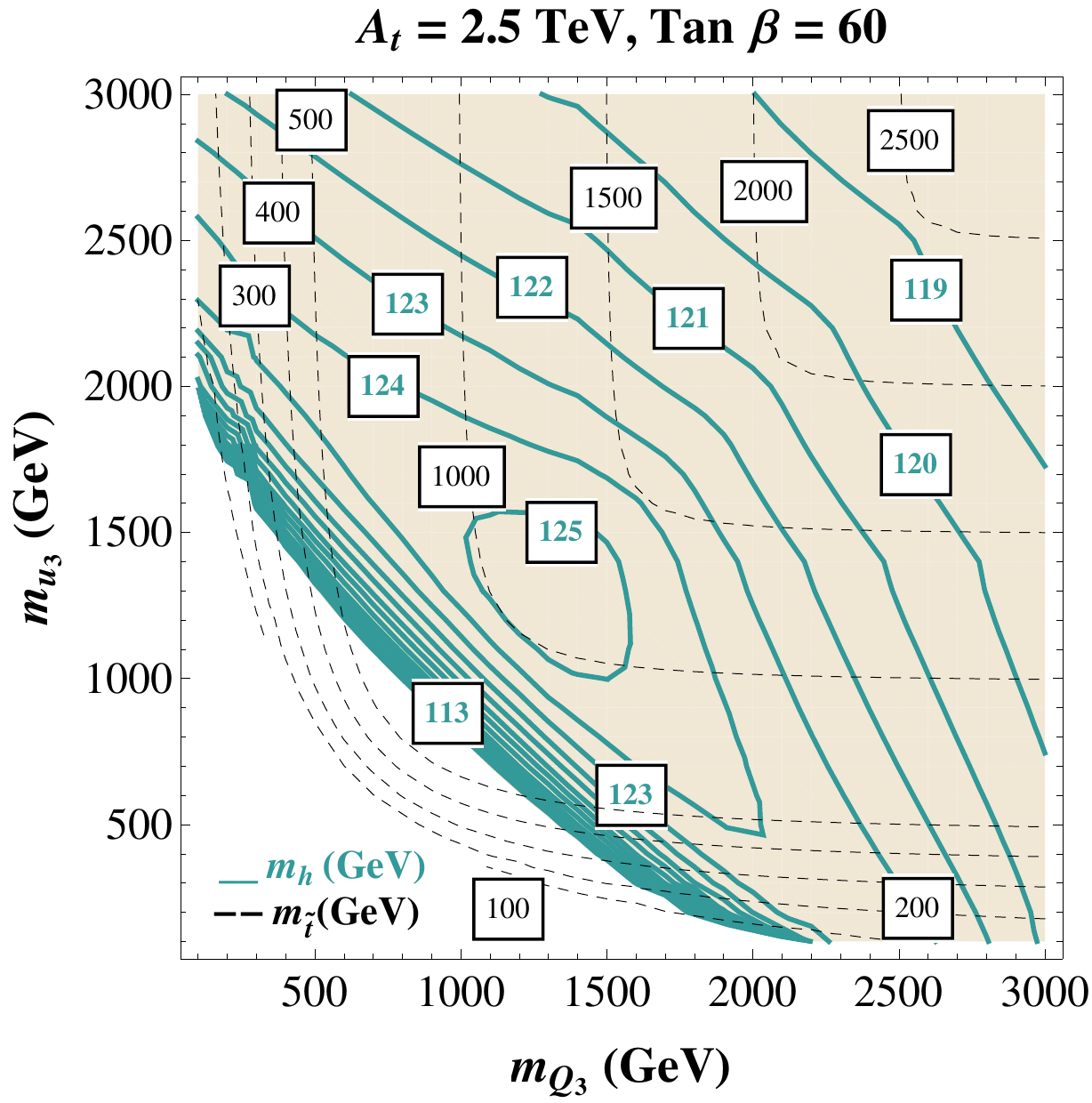} \\
\end{tabular}
\caption{Contour plots of the Higgs mass in the $m_{Q_3}$--$m_{u_3}$ plane, for different values of $A_t$ and $\tan\beta$. The stau soft masses have been fixed at $m_{L_3}^2 = m_{e_3}^2 = (350 \; {\rm GeV})^2$, while $\mu = 1030$ GeV and $A_\tau = 500$~GeV, leading to a lightest stau mass of about 135~GeV for $\tan\beta = 60$. The lightest stop masses are overlaid in dashed black lines.}
\label{Fig:countormhmq3mu3at246}
\end{center}
\end{figure}

\begin{figure}
\center
\begin{center}
\begin{tabular}{cc}
(a)&(b)\\
\includegraphics[width=0.5\textwidth]{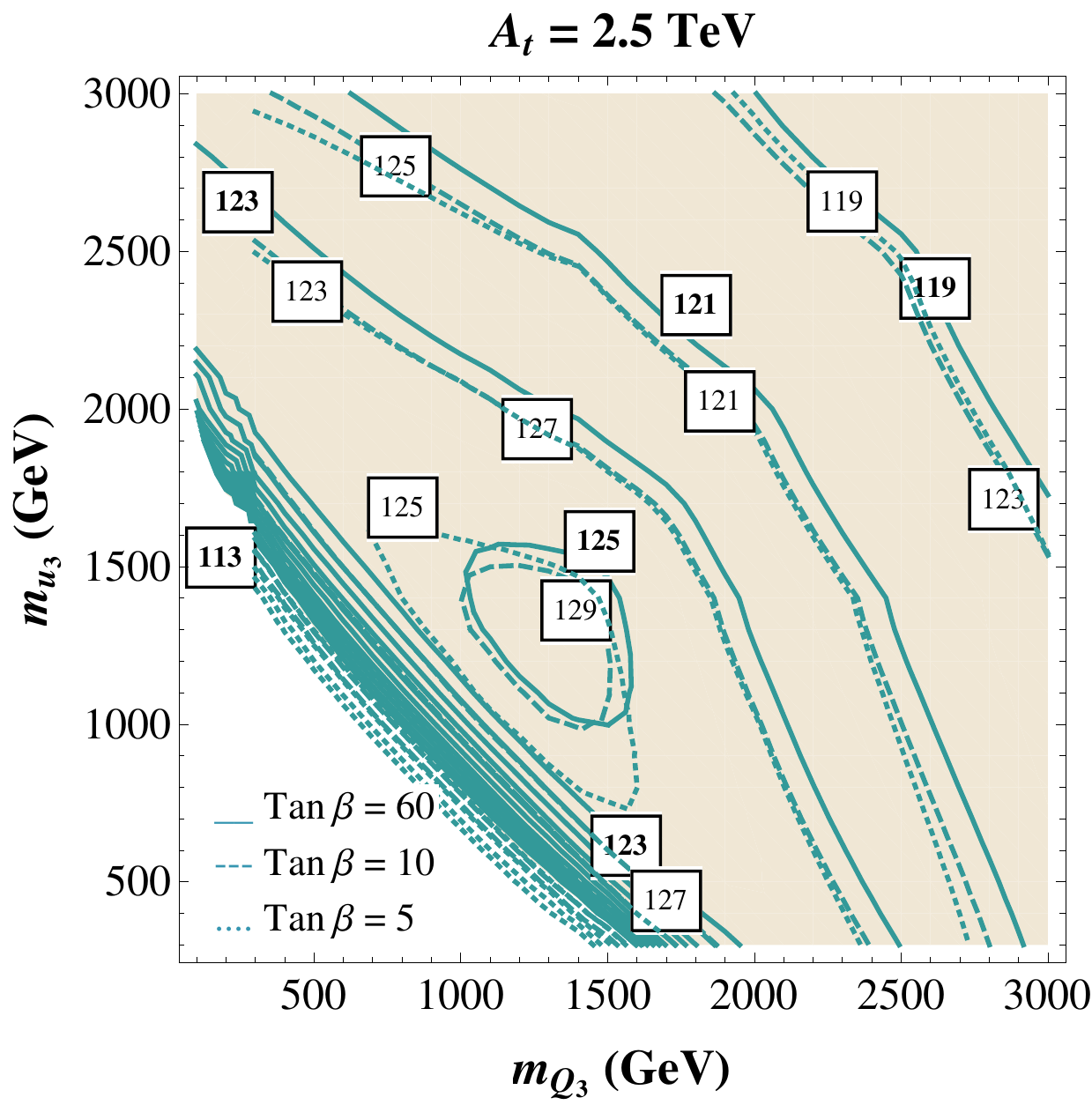} &
\includegraphics[width=0.5\textwidth]{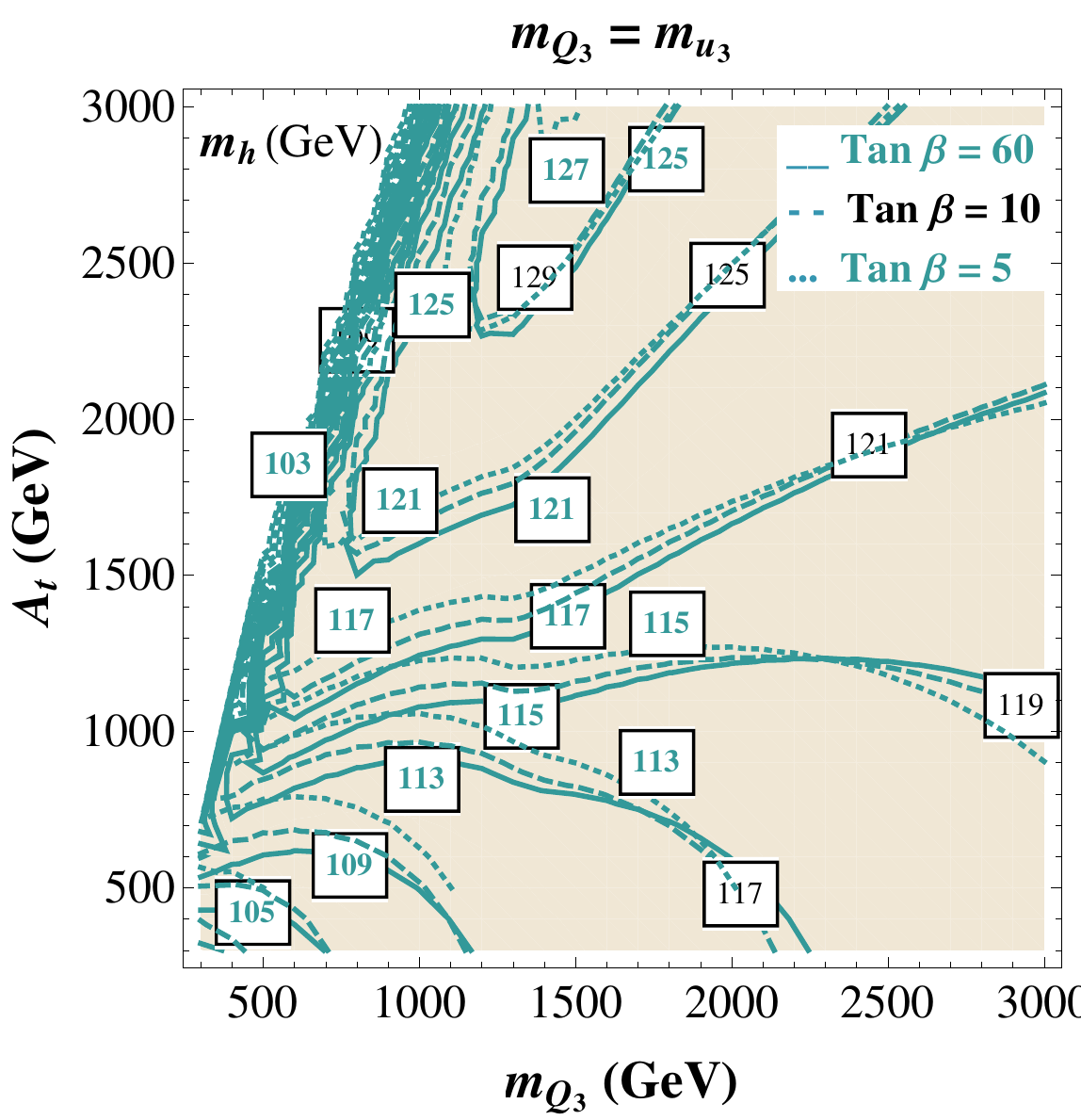}
\end{tabular}
\caption{Contour plots of the Higgs mass in the plane of soft supersymmetry breaking parameters in the stop sector. In (a), we show the Higgs masses for $A_t=2.5$ TeV for three different values of $\tan\beta$, $\tan\beta = 5$ (dotted lines, green~(grey) labels), $\tan\beta = 10$ (dashed lines, black labels) and $\tan\beta = 60$ (solid lines, green~(grey) labels). The masses for $\tan\beta = 60$ shown are smaller than the ones for $\tan\beta = 10$ mostly due to the negative effects from the staus (see Eq.~\ref{eq:mhstau}), and closer to the $\tan\beta = 5$ ones. In (b), the Higgs mass contours are shown for $m_{Q_3} = m_{u_3}$, varying the stop mixing parameter $A_t$. The stau supersymmetry breaking parameters have been kept at $m_{L_3}^2 = m_{e_3}^2 = (350 {\rm GeV})^2$ and $A_\tau = 500$~GeV, while $\mu = 1030$ GeV.}
\label{Fig:countormhmq3At}
\end{center}
\end{figure}

Results for the Higgs masses for different values of the stop mass parameters in the on-shell scheme and $\tan\beta$ in the $\bar{DR}$ scheme are shown in Figs.~~\ref{Fig:countormhmq3mu3at246}--\ref{Fig:countormhmq3At}. Throughout this paper we fix the gluino, wino and bino masses to 1.2 TeV, 300 GeV and 100 GeV, respectively. The right-handed down squark masses are fixed to $m_{d_i}=2$ TeV. In Fig.~\ref{Fig:countormhmq3mu3at246} we show the Higgs mass predictions (solid line contours) together with the stop mass predictions (black dashed contours) for different values of $A_t$ and $\tan\beta$ as a function of the soft SUSY breaking stop mass parameters, $m_{Q_3}$ and $m_{u_3}$. In the case of significant splitting of the stop soft masses, the mass of the heaviest stop is of the order of the largest soft stop mass, and as can be seen from Fig.~~\ref{Fig:countormhmq3mu3at246}, the mass of the lightest stop can be as low as $\sim 100$ GeV. This shows that the mass of the Higgs does not imply a hard lower bound on the squark masses. A lower bound for the squark masses will be determined by direct experimental searches. Note that, values of $A_t$ larger than $\sim 2$ TeV~\footnote{These large mixing parameters may only be avoided for very large values of the heaviest stop mass~\cite{Carena:2008rt}.} are required to achieve values of the Higgs mass in the region of interest.

In Fig.~\ref{Fig:countormhmq3At}~(a), we show the Higgs mass predictions for $A_t = 2.5$~TeV and three different values of $\tan\beta$ in the $m_{Q_3}$--$m_{u_3}$ plane. We observe that for similar values $m_{Q_3}$ and $m_{u_3}$ very large values of $\tan\beta$ ($=60$) lead to smaller values of the Higgs mass, if compared to sizable values of $\tan\beta$ ($=10$). This is due to the slepton contributions to the Higgs mass~(in the plot we are fixing $m_{L_3}^2 = m_{e_3}^2 = (350 \; {\rm GeV})^2$). Note that such large values of $\tan\beta$ are allowed since we are fixing large values of $m_A$ ($\sim 1000$ GeV) to satisfy LHC bounds from the non-standard Higgs boson search $H\rightarrow\tau\tau$~\cite{Chatrchyan:2011nx,atlastautau,cmstautau}.

In Fig.~\ref{Fig:countormhmq3At}~(b), we give contour plots of the Higgs mass in the $m_{Q_3}$--$A_t$ plane for different values of $\tan\beta$, assuming $m_{Q_3} = m_{u_3}$. We observe that in the case of no splitting between the two stop soft masses, values of $A_t$ above $\sim 1.5$ TeV are needed to achieve Higgs masses in the region of interest. In this case the mass of the lightest stop is naturally above a few hundred GeV.

\section{Production Rate of Higgs Decay into Photons}

The production rate of two photons associated with a SM-like Higgs decay may be increased by either increasing the gluon fusion production rate or by increasing the Higgs branching ratio into photons. Modifications of these rates may come from mixing effects or from extra particles running in the loops. We discuss these possibilities below.

\subsection{Mixing Effects}

The mixing in the Higgs sector can have relevant effects on the production rates and decay branching ratios. Mixing effects become particularly relevant for small values of the non-standard Higgs masses, $m_A$. It is known, however, that in most regions of parameter space, the mixing effects conspire to enhance the bottom decay width, leading to a suppression of the total production of photons and gauge bosons~(see, for instance Refs.~\cite{arXiv:1107.4354},\cite{Cao:2011pg}). However, the mixing in the Higgs sector may be modified for large values of the mixing parameters in the sfermion sector~\cite{Carena:1999bh}. Both stops, sbottoms and sleptons may have a relevant impact on the Higgs branching ratios. A suppression of the bottom decay width through mixing effects may have important consequences for the decay branching ratios of all the gauge boson decay channels.

Let us clarify the mixing effects in the CP-even Higgs sector. The mass matrix is given approximately by
\begin{equation}
{\cal M}_H^2  =
\left[
\begin{tabular}{c c}
$m_A^2 \sin^2\beta + M_Z^2 \cos^2\beta$          &  $-(m_A^2 + M_Z^2) \sin\beta \cos\beta +{\rm Loop}_{12}$ \\
$ -(m_A^2 + M_Z^2) \sin\beta \cos\beta +{\rm Loop}_{12}$         & $m_A^2 \cos^2\beta + M_Z^2 \sin^2\beta +{\rm Loop}_{22}$
\end{tabular}
\right]\;,
\end{equation}
where we have included the dominant mixing effects. The loop effects ${\rm Loop}_{22}$ are the loop corrections appearing in the second term of Eq.~\ref{mhsm}, divided by $\sin^2\beta$. Indeed, in the decoupling limit for large $m_A^2$,
\begin{equation}
m_h^2 \simeq
\left( {\cal M}_H^2 \right)_{11} \cos^2\beta +
\left( {\cal M}_H^2 \right)_{12} \cos\beta \sin\beta +
\left( {\cal M}_H^2 \right)_{21} \sin\beta \cos\beta +
\left( {\cal M}_H^2 \right)_{22} \sin^2\beta
\end{equation}
which reduces to Eq.~(\ref{mhsm}).

The loop-corrections to the $\left( {\cal M}_H^2 \right)_{12}$ matrix element are given approximately by~\cite{hep-ph/9808312,Carena:1999bh},
\begin{equation}
{\rm Loop}_{12}  = \frac{m_t^4 }{16 \pi^2 v^2 \sin^2\beta} \frac{\mu \tilde{A}_t}{M_{\rm SUSY}^2}  \left[  \frac{A_t \tilde{A}_t}{M_{\rm SUSY}^2} - 6 \right]
+ \frac{h_b^4 v^2}{16 \pi^2} \sin^2\beta \frac{\mu^3 A_b}{M_{\rm SUSY}^4}  + \frac{h_{\tau}^4 v^2}{48 \pi^2} \sin^2\beta \frac{\mu^3 A_{\tau}}{M_{\tilde{\tau}}^4}\,.
\end{equation}

The mixing in the CP-even Higgs sector may be now determined by
\begin{eqnarray}
\sin(2\alpha) =    \frac{2 \left( {\cal M}_H^2 \right)_{12}}{\sqrt{Tr[{\cal M}_H^2]^2 - det[{\cal M}_H^2] }}\;,
\nonumber\\
\cos(2\alpha) = \frac{\left( {\cal M}_H^2 \right)_{11}-\left( {\cal M}_H^2 \right)_{22}}{\sqrt{Tr[{\cal M}_H^2]^2 - det[{\cal M}_H^2] }}
\label{Eq:Mixing}
\end{eqnarray}
which reduce to $-\sin2\beta$ and $-\cos2\beta$ respectively, in the large $m_A$ limit. The convention is such that $0\leq \beta \leq \pi/2$~(although generically values of $\beta > \pi/4$ are considered), while $-\pi/2 \leq \alpha \leq \pi/2$, and, in the large $m_A$ limit, $\alpha = -\pi/2+\beta$.

The ratio of the tree-level couplings of the Higgs to $W$ bosons, top and bottom-quarks with respect to the SM ones are approximately given by
\begin{eqnarray}
hWW &  :  &\;\; \sin(\beta-\alpha)\,,
\nonumber\\
h t \bar{t} & : & \;\; \frac{\cos\alpha}{\sin\beta}\,,
\nonumber\\
h b \bar{b} & : & \;\; -\frac{\sin\alpha}{\cos\beta}\left[ 1 - \frac{\Delta h_b \tan\beta}{1 + \Delta h_b \tan\beta} \left( 1 + \frac{1}{\tan\alpha \tan\beta} \right)\right]\,.
\end{eqnarray}
As seen above, the coupling to bottom quarks is also affected by the $\Delta h_b$ corrections~\cite{hep-ph/9808312,Carena:2002qg}, which, however, do not modify the overall dependence of the bottom quark coupling on the mixing in the Higgs sector.

For moderate values of $\tan\beta$ and $m_A$, the loop effects are small and $\sin\alpha$ is small and negative while
\begin{equation}
 |\sin(2\alpha)| > | \sin(2\beta)|\;.
\end{equation}
Since $\cos\alpha \simeq \sin\beta \simeq 1$, this implies that $|\sin\alpha| >  \cos\beta$, leading to an enhancement of the bottom quark width which in turn leads to a suppression of the dominant SM Higgs decay branching ratios at the LHC. The couplings to top and $W$ bosons are not modified in this regime, but there is also a small decrease of the gluon fusion rate induced by the bottom-quark loop effects that have the opposite sign as the top quark loops and become enhanced in this regime.

For large values of $\tan\beta$ and moderate values of $m_A^2$, the values of $\sin\alpha$ tend to be very small, of order $\cos\beta$. A decrease of the bottom quark coupling can be obtained, for instance, if $|\sin(2\alpha)| < (\sin2\beta)$, which can be obtained by making the loop corrections ${\rm Loop}_{12}$ positive and sizable. Since the tree-level contribution for $({\cal M}_H^2)_{12}$ is suppressed by $1/\tan\beta$, the loop-corrections may be significant in the large $\tan\beta$ regime. It is well known that a suppression of the Higgs mixing can be achieved for large values of $\mu A_t <0$ ($\mu A_t >0$) for $A_t < \sqrt{6} M_{\rm SUSY}$ ($A_t > \sqrt{6} M_{\rm SUSY})$, as follows from Eq.~(\ref{Eq:Mixing}). Sizable values of $A_t$ are necessary to achieve a large modification of the Higgs mixing, what leads to values of the Higgs mass of about 120--125 GeV for stops masses of about 1~TeV. A benchmark scenario for Higgs searches at hadron colliders, named the ``small $\alpha_{\rm eff}$ scenario'', has been constructed due to this property~\cite{Carena:2002qg}. Large values of $\mu^3 A_{b,\tau} > 0$ may also lead to a significant effect for very large values of $\tan\beta$. Let us stress again that the overall effect of a suppression of the bottom quark width is an enhancement of not only the photon decay rate, but also of the $WW$ and $ZZ$ rates. A large suppression of the bottom-quark width, however, demands small values of $m_A$ and large $\tan\beta$, which are disfavored~\cite{arXiv:1107.4354} by the search for non-standard Higgs bosons at the LHC $H\rightarrow\tau\tau$~\cite{Chatrchyan:2011nx}--\cite{cmstautau}.  For instance, only a narrow region of the small $\alpha_{eff}$ scenario, for moderate $\tan\beta$ and $m_A \simeq 100$~GeV, for which the heaviest CP-even Higgs has SM-properties with a reduced bottom decay width, seems to survive these constraints~\cite{arXiv:1107.4354}.

\subsection{Light Stop and Sbottom Effects}

The Higgs decay rate into photons is induced by loops of charged particles. In the SM the main contribution comes from $W$ bosons and is partially suppressed by the contribution of the top quarks, which provides the second most important
contribution to the Higgs to $\gamma \gamma$ amplitude. In the MSSM, one can imagine that the presence of light sbottoms or light stops would contribute to this amplitude, and indeed, for sufficiently light squarks the decay branching ratio of the Higgs to $\gamma\gamma$ may be enhanced in certain regions of parameter space~\cite{Djouadi:1998az}.

\begin{figure}
\center
\begin{center}
\includegraphics[width=0.5\textwidth]{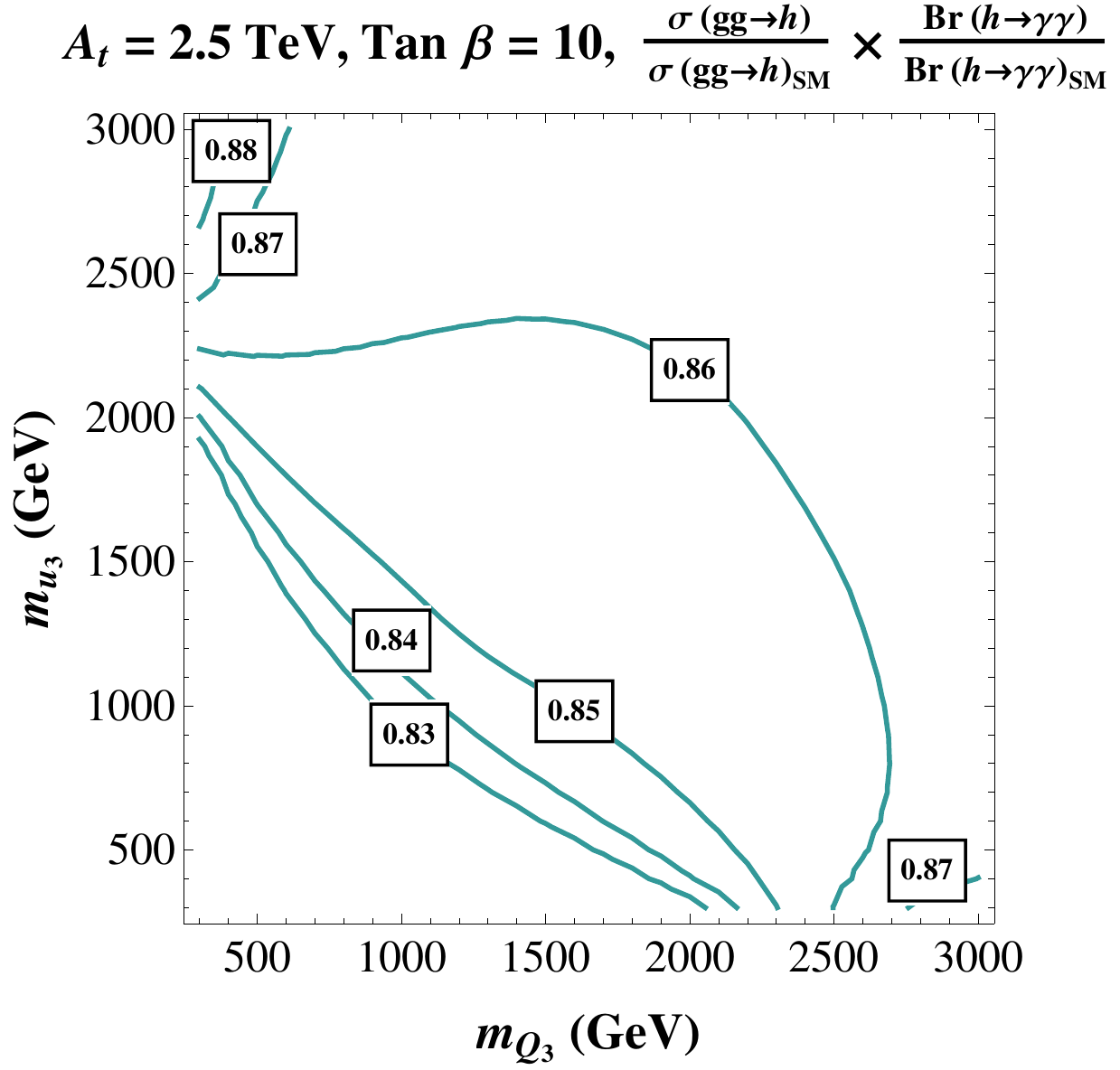}
\caption{Contour plots of the ratio of  the $\sigma(gg \to h) \times $ BR($h \to \gamma \gamma$) to its SM value, in the $m_{Q_3}$--$m_{u_3}$ plane, for $\mu = 1030$~GeV.}
\label{Fig:PhotonStop}
\end{center}
\end{figure}

Light squarks, with large mixing, can increase the photon decay branching ratio but, in general, this effect is overcompensated by a large suppression of the gluon fusion production rate~\cite{Dermisek:2007fi}, as it is shown in Fig.~\ref{Fig:PhotonStop}~\footnote{We found a small discrepancy between the  FeynHiggs and CPsuperH results for the branching ratio of the Higgs decay into photons in the limit of heavy sleptons, with FeynHiggs giving almost 10\% lower values than the SM even for heavy squarks. Since our computations were performed with FeynHiggs, the rates can be slightly higher than the ones shown in the figure.}. We find that, in general, for heavy third generation sleptons, for the region of third generation squark masses consistent with a 125~GeV Higgs the squark effects lead to a Higgs gluon fusion production times photon decay rate of the order of, or slightly lower than in the SM.

\subsection{Light Stau Effects}

A positive contribution to the $\gamma\gamma$ production rate, without modifying the gluon fusion rate, may only be due to loops of sleptons and charginos. In most cases slepton and chargino contributions tend to suppress the $\gamma\gamma$ rate. An exception to this rule is staus, in the presence of large mixing, which tend to enhance it.

Large mixing in the stau sector may be achieved for large values of $\mu$ and $\tan\beta$. This is due to the fact that the stau mass matrix is given by
\begin{eqnarray}
\mathcal M_{\tilde{\tau}}^2
\simeq
\left[
\begin{tabular}{c c}
$ m_{L_3}^2 + m_{\tau}^2 + D_L$  &
$ h_{\tau}v(A_\tau\cos\beta - \mu \sin\beta) $ \\
$ h_{\tau}v(A_\tau\cos\beta- \mu \sin\beta)$ &
$ m_{E_3}^2 + m_{\tau}^2 + D_R $\\
\end{tabular}
\right]
\end{eqnarray}
where $D_L$ and $D_R$ are the D-term contributions to the slepton masses~\cite{reviews}. Another condition that must be fulfilled is that the lightest stau is rather light, with a mass close to the LEP limit. For instance, for a value of $m_{L_3}^2 \simeq m_{e_3}^2 \simeq$ (350 GeV)$^2$, $A_\tau \simeq 500$~GeV, these conditions may be achieved
for $\mu \simeq 1$ TeV and $\tan\beta \simeq 60$. For these values
\begin{equation}
{\rm BR}(h \to \gamma \gamma) \simeq 1. 5 \; {\rm  BR}(h \to \gamma \gamma)^{\rm SM}
\end{equation}
may be obtained, together with no relevant effects in the Higgs gluon fusion production rate.

The dependence of $\sigma(gg \to h) \times BR(h\to \gamma\gamma)$ in the $m_{L_3}$--$m_{E_3}$ parameter space, for $\mu = 1030$~GeV, $A_\tau = 500$~GeV, as well as in the  $m_{L_3}$--$\mu$ parameter space for $m_{L_3} = m_{e_3}$ is shown in Fig.~4. Solid lines represent contours of equal photon rate, normalized to the SM value. Dashed lines represent contours of equal values of the lightest slepton mass. The squark sector was fixed at $m_{Q_3} = m_{u_3} = 2$~TeV and $A_t = 2$~TeV for $\tan\beta = 10$ and $m_{Q_3} = m_{u_3} = 1.5~TeV$, $A_t = 2.5$~TeV, for $\tan\beta = 60$, consistent with a Higgs mass of about 125~GeV. We have checked, however, that the results are insensitive to the exact value of the Higgs mass in the 123~GeV--127~GeV range. For $\tan\beta = 10$~(top panels in the figure) the stau mixing is small and no enhancement is observed in the total photon rate associated with Higgs production. On the contrary, for large values of $\tan\beta$~(bottom panels in the figure), for which the mixing is relevant, a clear enhancement is observed in the region of parameters leading to light staus, close to the LEP limit. As emphasized above, enhancements of the order of $50\%$ in the total photon rate production may be observed. The production rate of weak gauge bosons, instead, as well as the branching ratio of the Higgs decay into bottom quarks, remain very close to the SM one.

Let us mention in closing that large values of $A_\tau$ and moderate values of $m_A$ can lead to a suppression of the width of the Higgs decay into bottom quark via Higgs mixing effects, Eq.~(\ref{Eq:Mixing}), and a subsequent enhancement of the photon and weak gauge boson production rates. For instance, for $\tan\beta = 60$, $A_\tau \simeq 1500$~GeV, $m_A \simeq 700$~GeV, $\mu = 1030$~GeV and $m_{e_3} = m_{L_3} = 340$~GeV, one obtains a lightest stau mass of order 106~GeV, and
\begin{eqnarray}
\frac{\sigma(gg \to h)}{\sigma(gg \to h)_{\rm SM}} \frac{{\rm{BR}}(h \to \gamma \gamma)}{{\rm{BR}}(h \to \gamma\gamma)_{\rm SM}}  & = &1.96
\nonumber\\
\frac{\sigma(gg \to h)}{\sigma(gg \to h)_{\rm SM}} \frac{{\rm{BR}}(h \to V V^*)}{{\rm{BR}}(h \to V V^*)_{\rm SM}}  & = & 1.25    \;\;\;\;\;\; ( V = W,Z)
\end{eqnarray}
while ${\rm{BR}}(h \to b \bar{b}) \simeq 0.8 \rm{BR}(h \to b \bar{b})_{\rm SM}$. The LHC and the Tevatron colliders will be able to test these possible variations of the Higgs production rates in the near future.

\begin{figure}
\label{gammastau}
\begin{center}
\begin{tabular}{c c}
\includegraphics[width=0.50\textwidth]{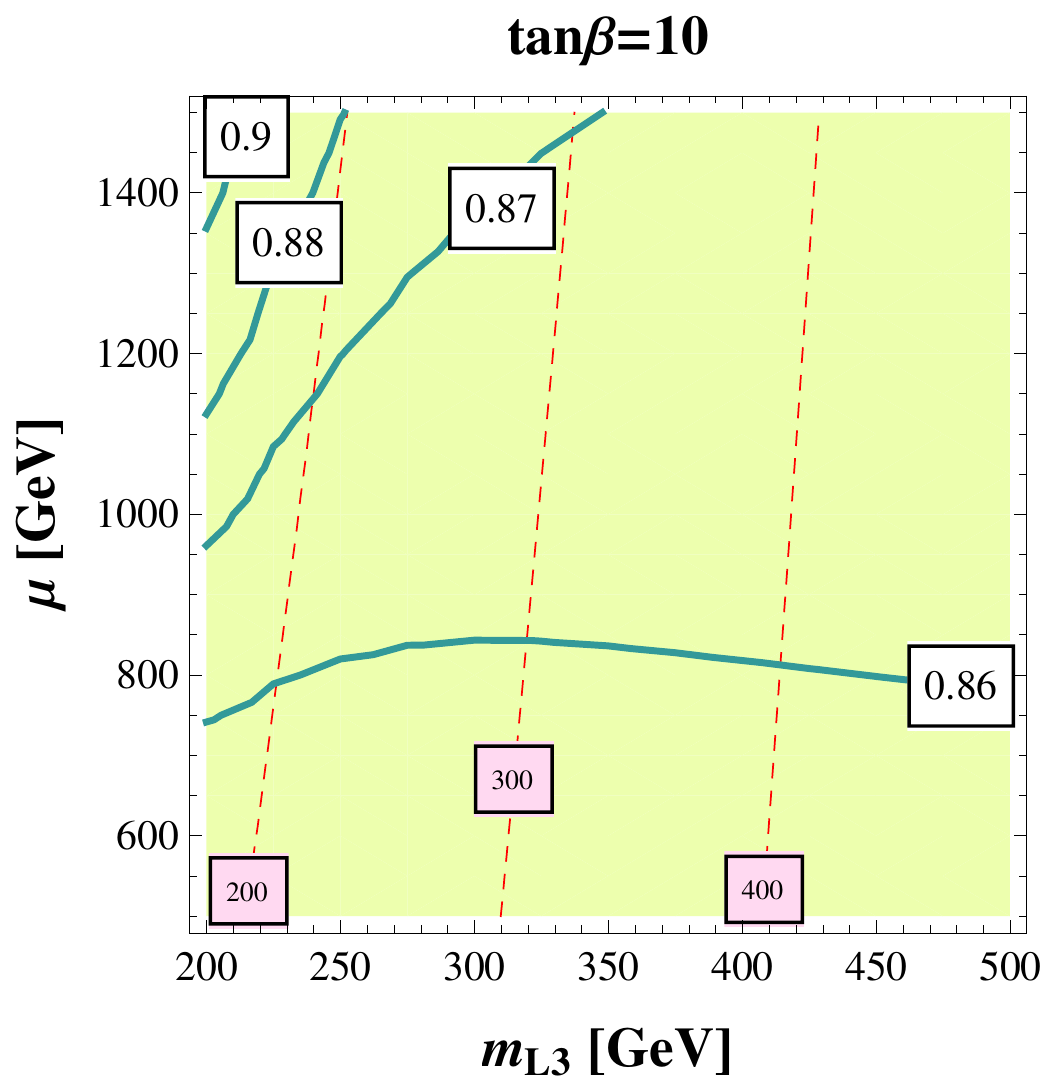}  &
\includegraphics[width=0.50\textwidth]{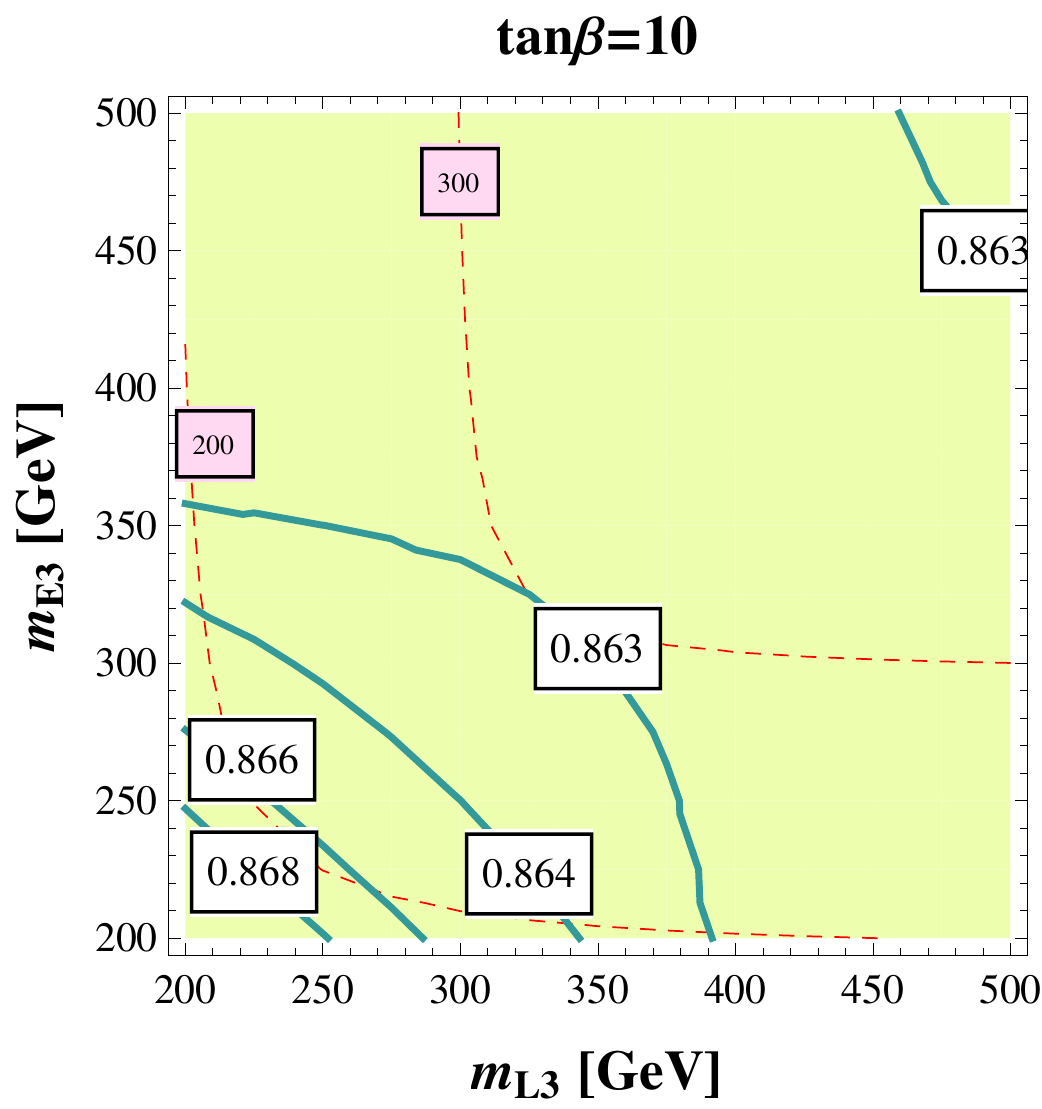}  \\
\includegraphics[width=0.50\textwidth]{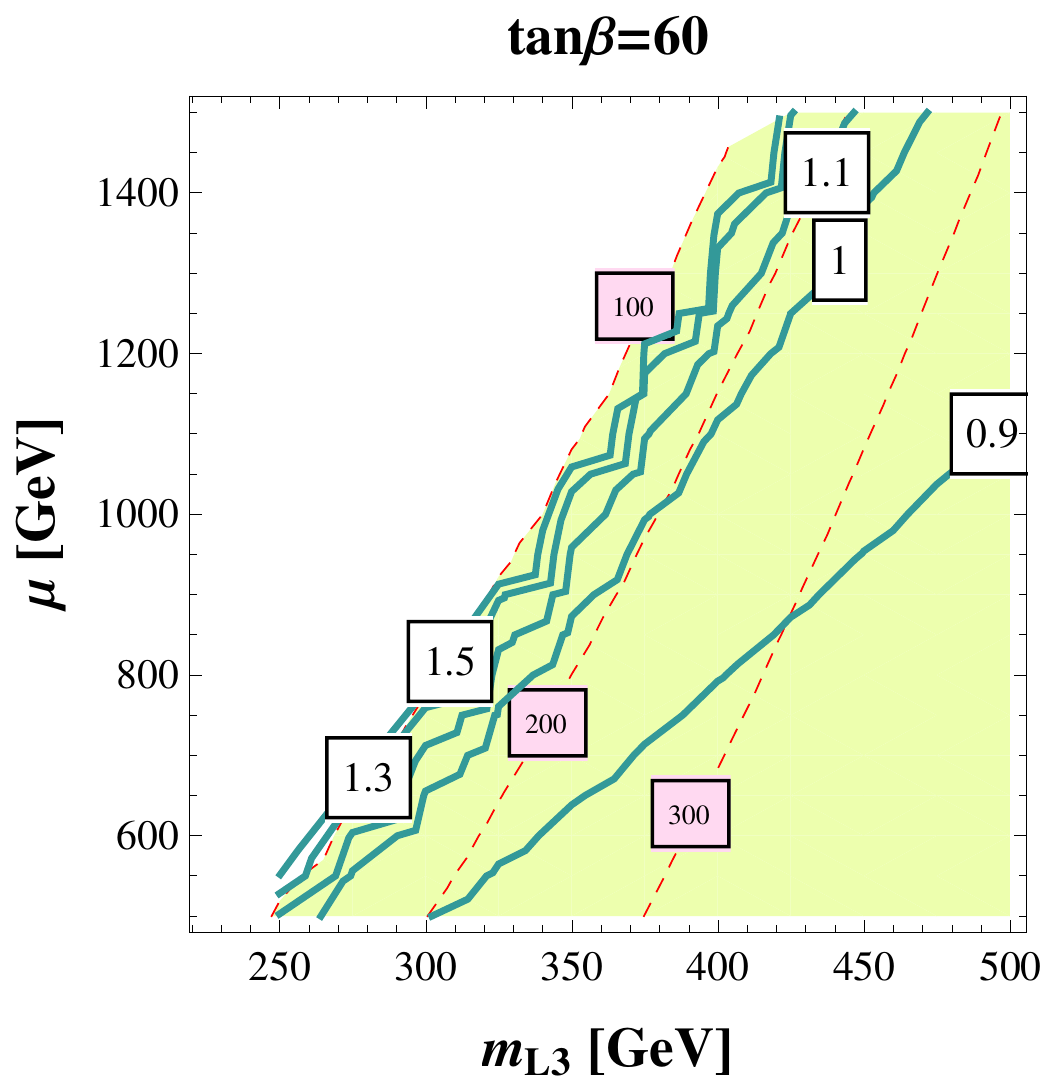}  &
\includegraphics[width=0.50\textwidth]{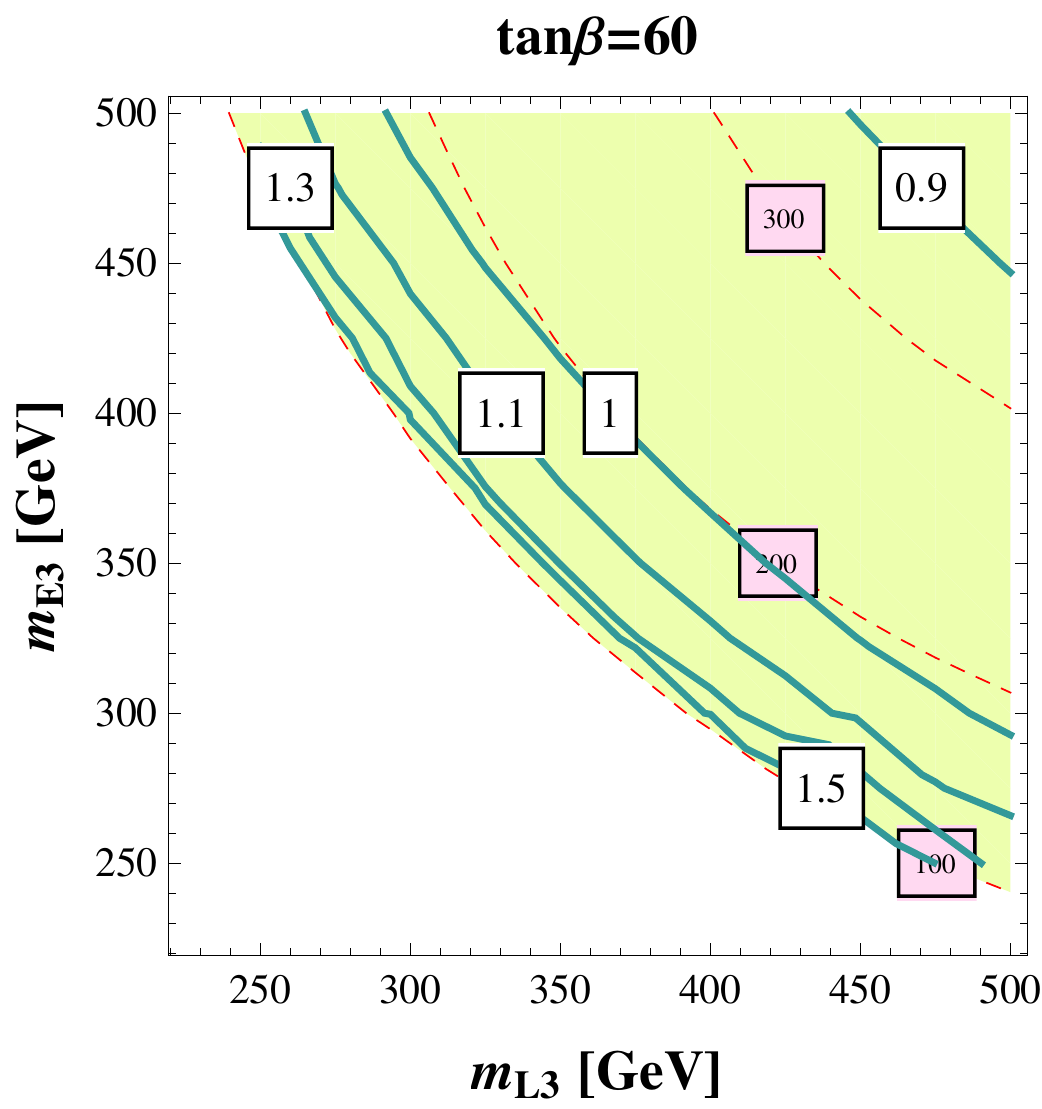}  \\
\end{tabular}
\end{center}
\caption{Contour plots of the ratio of  the $\sigma(gg \to h) \times $ BR($h \to \gamma \gamma$) to its SM value, in the $m_{e_3}$--$m_{L_3}$ plane, for $\mu = 1030$~GeV, as well as in the $\mu$ --$m_{L_{3}}$ plane, for $m_{e_3} = m_{L_3}$, and $\tan\beta = 10$ (above) and  $\tan\beta = 60$ (below). The red dashed lines are the contours at equal lightest stau masses. The yellow shaded area is the area satisfying the LEP bound on the lightest stau mass. Enhanced branching ratios are obtained for values of $\mu$ for which the stau mixing becomes relevant and the lightest stau mass is close to its experimental limit, of about 100~GeV. }
\end{figure}

\section{Conclusions}

The MSSM provides a well motivated extension of the SM, in which for a supersymmetric spectrum of the order of 1~TeV, the SM-Higgs mass remains below 130~GeV. Recent results from the LHC are consistent with the presence of a SM Higgs with a mass of about 125~GeV and a photon production rate that is similar or slightly larger than the SM one. This Higgs mass range is consistent with the presence of stops in the several hundred GeV range and a sizable mixing parameter $A_t \geq 1$~TeV. Lighter stops may also be obtained for relatively large values of the heaviest stop masses and sizable mixing parameters.

In general, for the stop mass parameters consistent with the 125~GeV Higgs, the gluon fusion rate tends to be slightly lower than in the SM. The photon decay branching ratio depends strongly on the CP-even Higgs mixing, which controls the bottom-quark decay width, and on the possible presence of light charged particles in the spectrum. Light squarks, with large mixing, required to get consistency with a relatively heavy Higgs mass of about 125 GeV, can increase the photon decay branching ratio but, in general, this effect is overcompensated by a large suppression of the gluon fusion production rate.

In this article, we have shown that light staus, with significant mixing, may strongly affect the photon decay rate if the lightest stau mass is close to the current experimental limit, of about 100~GeV. We have shown that di-photon production rates induced via Higgs production can be fifty percent larger than in the SM for a squark spectrum consistent with a 125~GeV Higgs mass. In general, large values of $\tan\beta$ and sizable values of $\mu$ are required to achieve these effects. Other experimental constraints, like the anomalous magnetic moment of the muon, flavor physics and the Dark Matter relic density would provide additional information to constrain the low energy soft supersymmetry parameters of the model. We reserve the study of these effects, as well as the associated LHC physics,  to a future work.

\subsubsection*{Note Added}
While this article was being completed several articles~\cite{Gogoladze:2011aa}--\cite{Arbey:2011aa} appeared in the literature which address the question of how to obtain a 125 GeV Higgs within low energy supersymmetric models. The effect of light third generation sleptons was not studied in these papers.

\subsubsection*{Acknowledgements}
Fermilab is operated by Fermi Research Alliance, LLC under Contract No. DE-AC02-07CH11359 with the U.S. Department of Energy. Work at ANL is supported in part by the U.S. Department of Energy~(DOE), Div.~of HEP, Contract DE-AC02-06CH11357. This work was supported in part by the DOE under Task TeV of contract DE-FGO2-96-ER40956.

\end{document}